%% file: main.tex
\documentclass[runningheads]{llncs}
\usepackage{graphicx}
\usepackage{subfig}
\usepackage{amsmath}
\usepackage{amssymb}

\begin{document}
	
\title{Workload Prediction of Business Processes
	\thanks{The authors would like to thank the CTI commission for the funding of this research project (project nb. 25500.2 PFES-ES).} 
}

\subtitle{An Approach Based on Process Mining and Recurrent Neural Networks}


\author{Fabrizio Albertetti \orcidID{0000-0001-6178-7378}  
	\and Hatem Ghorbel
}


\institute{Data Analytics Group, Haute Ecole Arc Ingénierie \\
	HES-SO, University of Applied Sciences Western Switzerland
}

\maketitle

\begin{abstract}
Recent advances in the interconnectedness and digitization of industrial machines, known as Industry 4.0, pave the way for new analytical techniques. Indeed, the availability and the richness of production-related data enables new data-driven methods.

In this paper, we propose a process mining approach augmented with artificial intelligence that (1) reconstructs the historical workload of a company and (2) predicts the workload using neural networks. Our method relies on logs, representing the history of business processes related to manufacturing. These logs are used to quantify the supply and demand and are fed into a recurrent neural network model to predict customer orders. The corresponding activities to fulfill these orders are then sampled from history with a replay mechanism, based on criteria such as trace frequency and activities similarity.

An evaluation and illustration of the method is performed on the administrative processes of Heraeus Materials SA. The workload prediction on a one-year test set achieves an MAPE score of 19\% for a one-week forecast. The case study suggests a reasonable accuracy and confirms that a good understanding of the historical workload combined to articulated predictions are of great help for supporting management decisions and can decrease costs with better resources planning on a medium-term level.

\end{abstract}

\input{corpus}

\section*{Acknowledgements}
The authors would like to thank the HMSA team and specifically Vincent Dessenne for their contribution.

\bibliographystyle{splncs04}
\bibliography{references-bpm}   


\end{document}

%% file: corpus.tex
\section{Introduction}
\label{sec:introduction}
The increasing use of information systems enables new analytics methods to support decision, especially for business process automation. Indeed, data-driven insights for optimizing processes, such as manufacturing, administrative, or logistics, to name only a few, proved to be valuable and contribute to cost reduction.

The problem of resources planning, that is to know what person is needed for what activity and for how long, specifically requires a good understanding of processes history. This problem can be approached by predicting the company's demand (i.e., the orders of the customers) and then predicting the company's supply (i.e., the workload, or the activities to conduct) to fulfill the demand.

To achieve this objective, this paper proposes a process mining approach augmented with machine learning. Our method is based on the reconstruction of the workload from a set of business process logs and uses a recurrent neural network to predict the customers' demand as a time series. Detailed activities of the workload for a medium term are then predicted with a replay mechanism from the history. 

The remainder of this paper is as follows: section \ref{sec:related} describes the background of the problem and relates similar work. Section \ref{sec:method} describes the proposed method. Section \ref{sec:experiment} illustrates the method with a use case, providing additional insights of the problematic. The use case serves also for the scoring of the workload prediction.

\section{Background and Related Work}
\label{sec:related}
Process mining is a research area that deals with discovery, verification of conformance and enhancement of processes based on the analysis of process traces.

In the literature, several methods have been proposed to address issues related to process enhancement, most of them still considered as ad hoc tasks \cite{Tiwari2008}. These enhancements include the prediction of the remaining time for a case, the prediction of the next activity in a trace of execution, or the prediction of a risk, to name only a few. \cite{Verenich2018} provide a survey of predictive methods for the monitoring of remaining time is and classify them into three categories, that is \textit{generative} methods, \textit{discriminative} methods, and \textit{hybrid} methods.

Discriminative methods include neural networks, which have been widely applied for enhancement tasks in process mining. More specifically, recurrent neural networks (RNN) are particularly adapted for predicting the next activity due to their capacity to handle sequences. In that perspective, \cite{Tax2016} compare different recurrent neural network architectures to predict the next activity and its timestamp. \cite{Hinkka2018} also use RNN, but to classify process instances with two boolean labels, related to the duration of the case and the need for more information to complete the case. Furthermore, they compare long-short term memory (LSTM) and gated recurrent units (GRU), which are two types of RNNs, and observe that while they hold similar accuracies, GRUs always have a shorter training time than LSTMs. In another study, \cite{Polato2018} predict a fixed number of activities for a running case. They represent alternatives with transition systems, each state trained with a different model, and predict a sequence. \cite{Tax2016} propose a model to predict the completion of a running case. Similarly, \cite{Ceci2014} predict the following activity of a trace and the completion time. 

Our approach differs slightly from these studies with these two observations: (a) in complement to predicting the completion of a running case, we also predict completely new cases that are triggered by the prediction of customer orders; and (b) the prediction of the completion of a running case is performed with a replay mechanism based on traces similarity and frequency. Because of these differences, we denote this research as \textit{workload} prediction. Indeed, as suggested in \cite{vanderAalst2012}, predicting the remaining flow time should consider not only the process instance, but also its context (i.e., the workload and resource availability).

\section{The Proposed Method}
\label{sec:method}
\begin{figure}[h]
	\centering
	\includegraphics[width=0.95\textwidth]{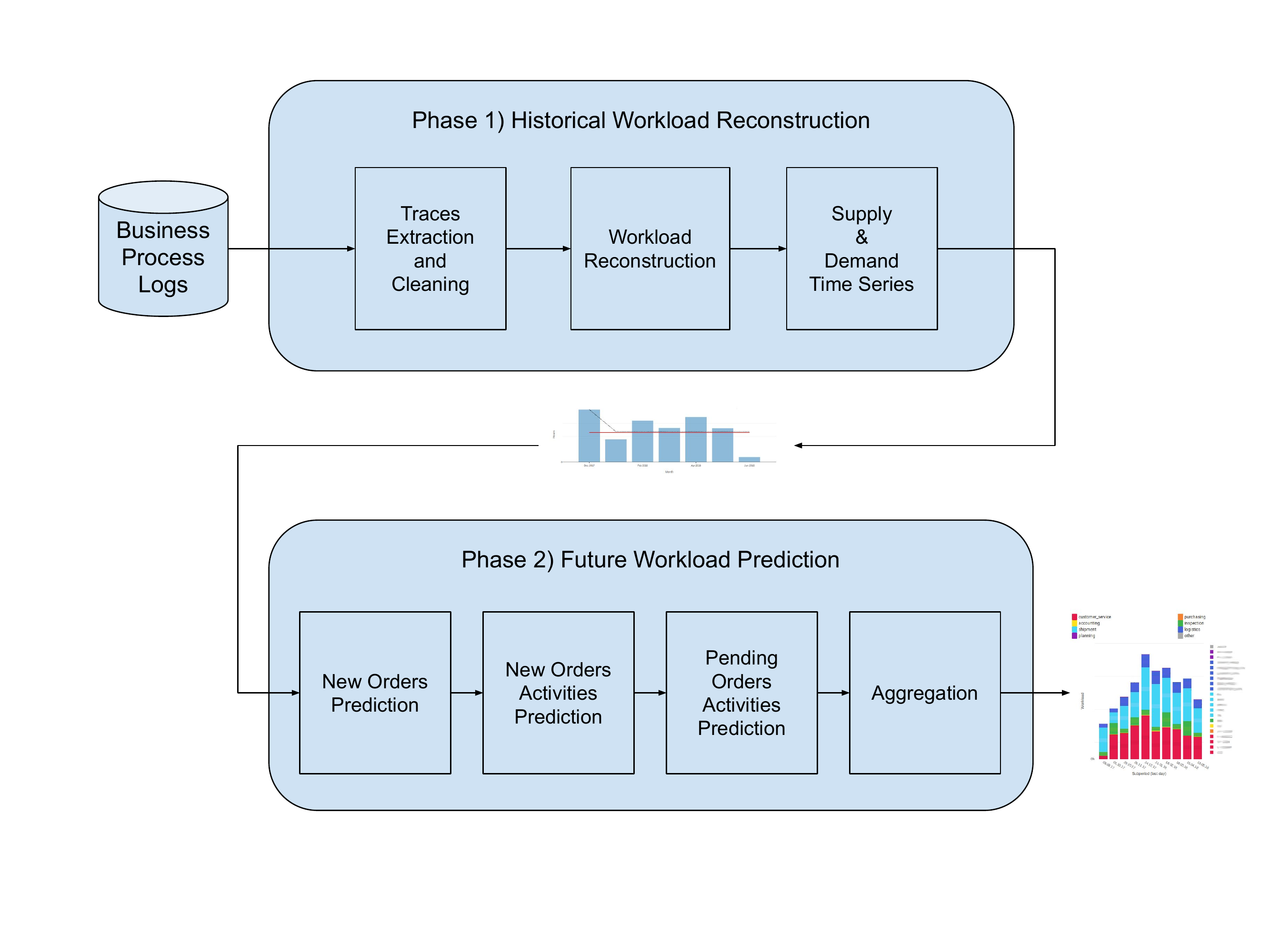}
	\caption{Overview of the proposed method}
	\label{fig:process}
\end{figure}

Our method consists of two phases (see Fig. \ref{fig:process}). The first phase is the reconstruction of the historical workload, i.e., the representation of the amount of work of each business unit aggregated on a daily basis. For that purpose, traces from business process logs are extracted and cleaned to represent the workload as time series.

The second phase is the prediction of the future workload and its corresponding activities. The workload prediction is based on the prediction of customer orders with a recurrent neural network. Based on these predicted quantities, the corresponding activities can be predicted with a replay mechanism (i.e., replaying the most frequent traces of similar orders from the history). Similarly, the activities of the running orders are completed. The final result is an aggregation of the new orders and the running orders, which represents the predicted workload and its underlying activities.

To apply the method on any company, the business process logs must contain the traces of a process, that is a sequence of activities identified by a business case. More specifically, the logs should:

\begin{itemize}
	\item represent a set of traces according to the process mining description of \cite{vanderAalst2016} (i.e., each activity being described by a start date, an end date or estimated duration, a name, a resource, and a case identifier);
	
	\item cover a period of several months (at least one year to obtain meaningful results);
	
	\item cover all the activities and business units of interest, to be able to fully reconstruct the investigated process;
	
	\item have a fine-grained granularity of the activities (i.e., a rich definition of activity types, a precise duration, and an attribution to resource and a business unit for the (non-)interchangeability).
\end{itemize}

Based on these assumptions, the workload is then predicted by applying the following five-steps method:

\begin{enumerate}
	\item \label{step:reconstruct} \textbf{Reconstruct the historical workload as time series.} The workload is an amount of work expressed on a daily basis. These quantities must be linked to a resource, a business unit, and a case identification (i.e., any field that can relate the work to a specific project or an order). 
	
	Another way to imagine the workload is as a function of the demand and the supply of the company. The demand can be modeled as the number of customer orders (i.e., the number of articles that needs to be produced) and the supply (i.e., the amount of work required to fulfill these orders).
	
	\item \textbf{Predict the new orders.} The prediction of the  orders is based on a machine learning model. The customer orders are represented as time series and can be aggregated by article type or used as multiple time series, one for each type. In the latter case, a model is trained for each type and the predictions are aggregated afterwards.
	
	Given a time series of K observations \( \mathbf{x_{t-K}^{t-1}} = x_{t-K}, x_{t-K+1}, ..., x_{t-1} \) representing historical orders, the objective is to predict the next observation \( x_{t} \). The prediction model is then denoted as the function \( f: \bbbr^{K} \mapsto \bbbr^{1} \):
	
	\begin{equation}
	f( \mathbf{x_{t-K}^{t-1}} ) = x_{t},
	\label{eq:prediction_function}
	\end{equation}
	
	where our prediction function $f$ is a trained recurrent neural network (RNN) with a dense layer for regression on top of it (see Fig. \ref{fig:rnn}). Each input time step of the RNN is fed by its respective observation from \( \mathbf{x_{t-K}^{t-1}} \) and some optional exogenous features, such as marketing promotions influencing the demand or economical indices of the customer's country.
	
	To predict a forecast window of more than one observation, we use the `re-injection' method by applying the same method on translated input window (i.e., the last predicted value \( x_t \) becomes the last element of our historical window and is fed for predicting \( x_{t+1} \) given \( \mathbf{x_{t-K+1}^{t}} \)). To deal with zero values and non-stationary time series, other transformations can be applied beforehand, such as exponential smoothing or differentiation.
	
	\begin{figure}[h]
		\centering
		\includegraphics[width=0.8\textwidth]{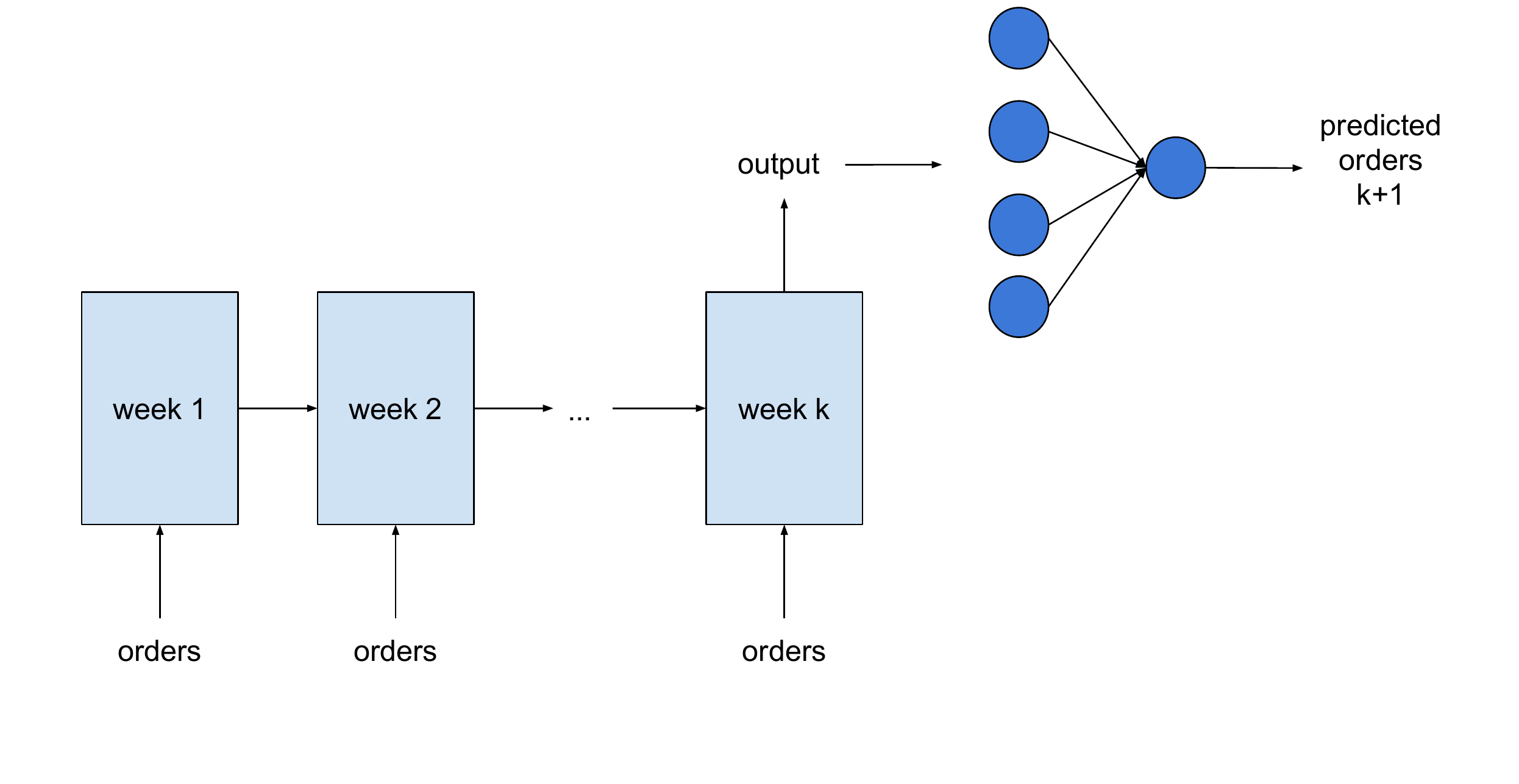}
		\caption{Architecture of the model for customer orders predictions. A dense layer is connected to the output of the last time step of the RNN for regressing the next value of the time series.}
		\label{fig:rnn}
	\end{figure}
	
	\item \textbf{Predict the activities of the new orders.} Since the predictions represent simple quantities of ordered articles, a more detailed representation of the workload is proposed. The objective is to translate customer orders into a set of related activities, i.e., traces, providing detailed information of the workload. These activities correspond to the diluted workload required to fulfill the predicted orders. This translation is based on `replaying' the most frequent similar orders of the historical workload.
	
	The similarity is based on the article type, that is all traces with the same article type are considered. To avoid replaying outliers, we select only 80\% of the most frequent traces. These traces are sampled in accordance to their distribution and are simply diluted over time with the same scale.
	
	\item \textbf{Predict the completion of the running orders.} To predict an exhaustive workload, we also include running orders, which were not considered in the previous step. Running orders are orders that have already started but have not been completed yet. They therefore do not depend on the new orders predicted.
	
	In this case, we use a slightly different replay mechanism than for the new orders activities from the previous step. For each running orders, we compare it to all historical traces of the same article type and compute a similarity based on the Levenshtein distance.
	
	A given order is represented by its trace as a string, i.e., a sequence of activities \( (a_1, a_2, ...) \) chronologically sorted. The Levenshtein distance is used to compute a distance between two traces \( O_i \) and \( O_j \) and is given by \\ \( lev_{O_i, O_j}(|O_i|, |O_j|) \), where:
	
	\[
	lev_{s, t}(m, n)=min
	\begin{cases}
	lev_{s, t}(m-1, n)+1 \\
	lev_{s, t}(m, n-1)+1 \\
	lev_{s, t}(m-1, n-1)+1_{(s_m \neq t_n)}
	\end{cases}
	\] 
	
	When $min(m, n)=0$, then
	\[
	lev_{s, t}(m, n)=max(m, n).
	\]
	
	\( 1_{(s_m \neq t_n)} \) is the indicator function equal to 1 when \( s_m \neq t_n \) and 0 otherwise, and \(lev_{s, t}(m, n) \) is the distance between the first $m$ characters of $s$ and the first $n$ characters of $t$.
	
	This distance intuitively corresponds to the minimum number of insertions, deletions, and modifications for transforming one trace into another.
	
	Given these distances, the a running case is aligned with the most similar trace, starting from the right-hand side of the string. The remainder of the similar trace is then replayed for completing the running case.
	
	\item \textbf{Predict the final workload activities.} The predicted final workload is the aggregation of two components: (a) the predicted activities of the new orders and (b), the predicted completions of the running orders.
\end{enumerate}

\section{Use Case: Heraeus Materials SA}
\label{sec:experiment}

For illustration and evaluation purposes, this Section details our method with the specific case of Heraeus Materials SA (HMSA), a leading industry manufacturer of high-value micro-components. This study focuses only on one of their sites in Switzerland.

Several years of their activities are stored within their information systems, of which 5 were considered usable for this use case. The main subject areas covered by these logs are customer orders, production orders, and articles. The traces were reconstructed from de-normalized spreadsheets extracted from their ERPs, containing a list of activities related to the following business units: customer service, logistics, purchasing, accounting, shipment, planning, and inspection. 

About 40 different ERP activity types have been stored since 2013, totaling around 2 million of events on 5 different files. All these activities can be grouped into 3000 customers orders totaling 8000 positions, 20'000 production orders, 100 customers, 31 article types, and more than 700 unique articles. The granularity chosen in this case study for the orders is at the article type level, that is the customer demand will be predicted for each of the 31 article types and the combined to form the overall demand. 

\textit{Please note that in the remainder of this section, many absolute quantities and activities names are not reported in the illustrations for confidentiality purposes.}

\subsection{Traces Extraction and Cleaning}
The extracted data for this research were simply de-normalized csv files, not fully exhaustive traces. Much work of pre-processing was therefore necessary to reconstruct the underlying workload, and a lot of time was spent to capture the many implicit business rules. We merged these de-normalized csv files and created two perspectives that can serve as a basis of any analytical task. One perspective regroups all necessary data for tracing logistics details related to any given customer order and the other perspective reconstructs logistics details for any given production order.

To automate the update of these perspectives upon arrival of new data, we created a fully automated transformation pipeline. Some tests were also included to ensure the integrity of the reconstruction.

\subsubsection{Business Process Analysis}
Based on the created perspectives, a process representation of the activities can be generated by simply importing the traces into some process mining visualization software.

HMSA already has some expertise with a process visualization tool for some part of their business intelligence activities. This tool supports process discovery and delivers flexible visualizations and descriptive statistics. Basically, it supports HMSA for a data-driven analysis of their historical data, limited to a specific process.

\begin{figure}[h]
	\centering
	\subfloat[All traces without filter]{
		\includegraphics[width = 0.22\textwidth]{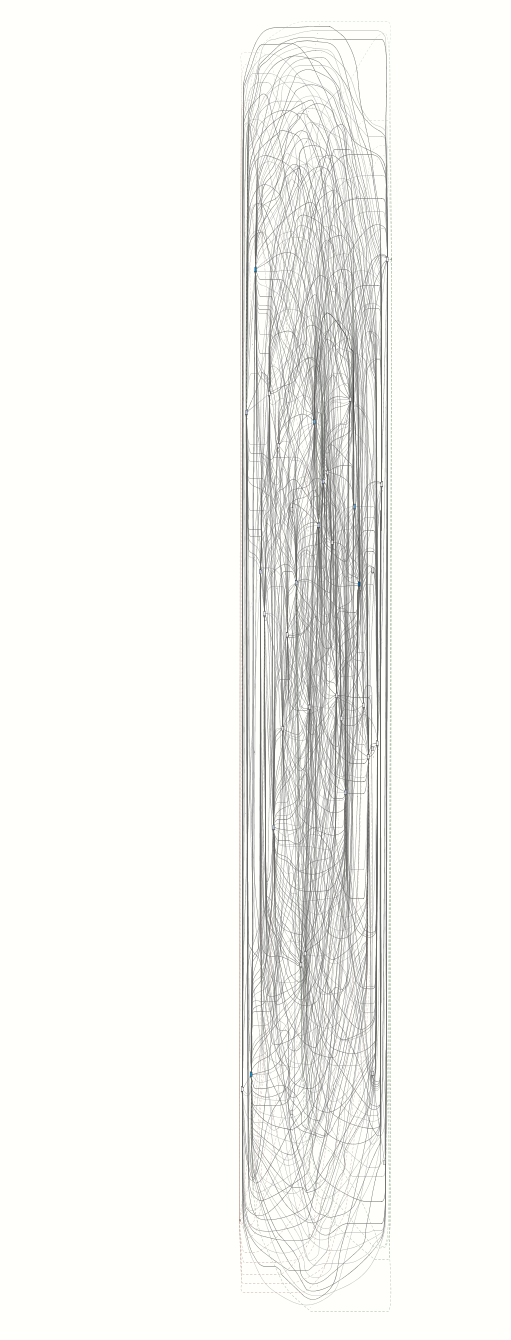}}
	\hspace{0.5cm}
	\subfloat[Most frequent paths only (1\%)]{
		\includegraphics[width=0.45\textwidth]{./process_map_orders_filter}}
	\caption{Fuzzy cognitive map of the customer-related orders activities}
	\label{fig:process-map}
\end{figure}

More broadly, their entire business process can be discovered from our reconstructed perspectives. Without filtering the traces, nothing but a ``spaghetti'' map representing a high variability of the activities flow can be understood. However, by setting a frequency filter on the activities or on the transitions, a more concise map, as in Figure \ref{fig:process-map}, can be delivered.

Several preliminary analyses have been performed from these perspectives, such as the comparison of production strategies. The objective of this latter was to quantify the use of two different production strategy given an article type for a specific period. A key indicator was created to represent the ratio of manufacturing starting (a) before the customer order (``make to stock'') and (b) after the customer order (``make to order'').

\subsection{Historical Workload Representation}
The historical workload is reconstructed as such. On one hand, the company demand is represented by daily time series of the customer orders, aggregated by article type. The observations being the number of positions of an order. An average smoothing on a 12-week triangle-shaped window was applied to each time series to reduce some part of the variance. On the other hand, the company supply is represented by all the activities related to these orders, that is the amount of work required to fulfill the demand. The observations here are the duration of these activities aggregated by article type.

\subsection{Prediction of the Workload}
As already stated in Section \ref{sec:method}, workload prediction is based on (a) the prediction of new orders and (b) the prediction of the completion of the running orders. The final result constitutes the set of activities for a forecast window, known as the workload. Predicting workload contributes to resources planning in two ways: it helps in the understanding of the medium-term demand and the required human workforce to fulfill that demand.

\begin{figure}[h]
	\centering
	\includegraphics[width=1\textwidth]{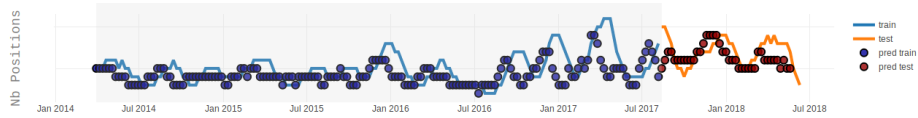}
	\caption{Prediction of the customer orders on a real test set with a prediction window of 41 weeks (based on a one-week forecast model with re-injected predictions) for a given article type. The predicted values are represented by red dots on the white area. The averaged MAPE propagated on these 41 weeks is of 47\%.}
	\label{fig:orders-prediction}
\end{figure}

For that purpose, the prediction function we used in place of Eq. \ref{eq:prediction_function} is a GRU-based recurrent network with a dense layer for regression. The learning window is made of 12 weeks (i.e., time steps) trained on an individual root mean squared error (RMSE) loss, on 100 epochs. For each of the 31 article types, a different model is trained on its corresponding training set (illustrated in Fig. \ref{fig:orders-prediction}). Each article type has a data set of 232 observations, of which 20\% dedicated for the test set. The hidden size of the GRU units are of 64, and a dropout of 0.2 has been set at the training phase to avoid overfitting. Some exogenous features were also included, such as a one-hot encoding of the date to capture day and month seasonality, the number of unique customers, and the number of unique countries. The intuition behind these exogenous features is that they implicitly capture a fraction of the demand variance (e.g., given that some article types are ordered only by very few customers, knowing this number may lead to more accurate results).

The evaluation for a single time step has an average RMSE of 0.72 over the 31 article types and a mean absolute percentage error (MAPE) of 19\%. The 0.72 RMSE means that on average the magnitude of the error (without sign) for one predicted week is of 0.72 orders. This score seems to be reasonable given that the objective is not to schedule production but to get insight of the human workforce required per business unit on a medium-term forecast. 

More specifically, most of the article types have stationary time series and seem to have a reasonable variance, but a few types present a difficult challenge. For instance, some weekly values are near to a random distribution, and therefore are very sensitive to the aggregation level. A high variance between two weeks is observed, even if the yearly quantities seem to follow a distribution. To avoid these substantial jumps, we use a centered exponential smoothing window averaged on 3 time steps.

It should be noted that for a prediction of more than one time step, a naive estimation of the mean error consists in multiplying the single time step error by the number of weeks. This estimator proves to be optimistic, because the error at one time step of the forecast window propagates to each subsequent time step. The bigger the forecast window is, the higher the error will be.

\begin{figure}[h]
	\centering
	\includegraphics[width = 0.45\textwidth]{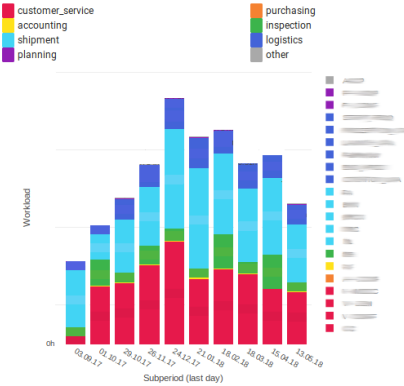}
	\caption{Prediction of the final workload activities (as a result of adding the predicted activities of the running orders and the new orders)}
	\label{fig:workload-prediction-final}
\end{figure}

To obtain the final workload, the remaining steps are applied as described in Sec.~\ref{sec:method}. The predicted new orders are used to generate their corresponding activities with a replay mechanism filtered on the 80\% most frequent traces. Then, the running orders are completed using another replay mechanism based on the Levenshtein distance. Finally, all the predicted activities are combined to represent the total workload, such as illustrated in Fig.~\ref{fig:workload-prediction-final}.

\subsection{Visualization Dashboards}
To further visualize and explore these results, we created a series of dashboards. Developed in Python with the Dash library from Plotly, they deliver insights and indicator about the past and the future. They also contribute to further explain the rationale of the predictions to the end users. These dashboards are categorized into three categories. 

\begin{figure}[h]
	\subfloat[Reconstruction of the historical workload represented as a time series]{
		\includegraphics[width = 0.45\textwidth]{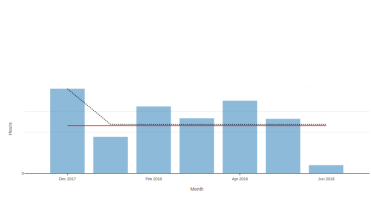}
		\label{fig:dashboard-workload-timeline}
	}
	\hspace{0.5cm}
	\subfloat[Prediction of the customer orders]{
		\includegraphics[width = 0.45\textwidth]{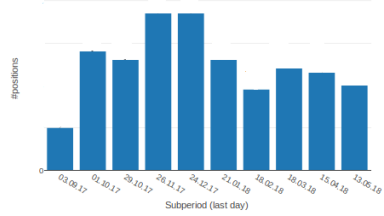}
		\label{fig:dashboard-prediction-orders}
	}
	\\
	\subfloat[Prediction of the customer orders activities]{
		\includegraphics[width = 0.45\textwidth]{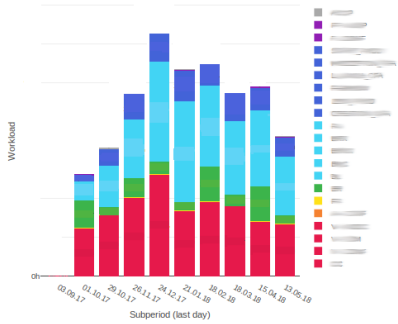}
		\label{fig:dashboard-prediction-activities}
	}
	\hspace{0.5cm}
	\subfloat[Prediction of the running orders completions]{
		\includegraphics[width = 0.45\textwidth]{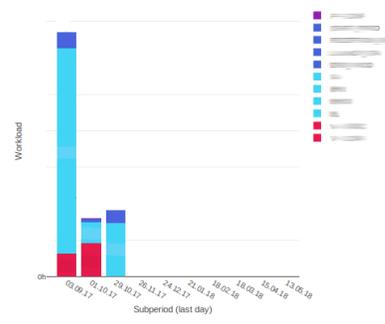}
		\label{fig:dashboard-prediction-running}
	}
	\caption{Overview of 4 of the developed dashboards.}
	\label{fig:dashboards}
\end{figure}

\paragraph{Descriptive dashboards} use descriptive statistics to understand and quantify past decisions:
\begin{itemize}
	\item the \textit{overview} dashboard (running work and key performance indicators for the current period);
	
	\item the \textit{process timeline} dashboard (a graphical view of all the process activities on a timeline, with an adjustable filter on any given article);
	
	\item the \textit{workload timeline} dashboard, illustrated in Fig. \ref{fig:dashboard-workload-timeline} (a reconstruction of the workload on a timeline, with adjustable filters); and 
	
	\item the \textit{stock timeline} dashboard (a graphical view on the history of the stock availability of every component).
\end{itemize} 

\paragraph{Exploratory dashboards} support the idea of simulations. They are not limited to descriptive statistics and explore new ideas or business questions with quantitative answers:

\begin{itemize}
	\item the \textit{components dependency graph} dashboard (a data-driven discovered taxonomy describing the components necessary to produce a given article type);
	
	\item the \textit{article type signature} dashboard (a representation of the most likely process realization of a given article type); and
	
	\item the \textit{shift capacity analysis} dashboard (an indicator of the minimum duration required to production of an order for a given article).
\end{itemize}

\paragraph{Predictive dashboards} have true analytic capabilities, that is they require supervised models for making prediction that learn from historical data:

\begin{itemize}
	\item the \textit{new orders prediction} dashboard, illustrated in Fig. \ref{fig:dashboard-prediction-orders} (the details of the predicted customer orders for a given forecast window and a given article type);
	
	\item the \textit{orders activities} dashboard, illustrated in Fig. \ref{fig:dashboard-prediction-activities} (the details of the corresponding activities from predicted orders);
	
	\item the \textit{running orders completion} dashboard, illustrated in Fig. \ref{fig:dashboard-prediction-running} (the details of the completion of running orders);
	
	\item the \textit{capacity prediction} dashboard (a prediction of the seasonality of workforce availability); and
	
	\item the \textit{activities workload prediction} dashboard, illustrated in Fig. \ref{fig:workload-prediction-final} (the final predicted amount of work and its corresponding activities required to fulfill all customer orders for a given forecast period).
\end{itemize}

\section{Conclusions and Future Work}
\label{sec:conclusion}
This paper proposes a data-driven method for medium-term human workload prediction. For that purpose, a representation of a company supply and demand is reconstructed from process logs and is fed as time series into a recurrent neural network. The predicted orders are then translated into a set of corresponding activities to form the predicted workload. Several dashboards are also describing as tools for supporting decisions.

The evaluation score of the method on a test-set is interpreted as reasonable for the given objective, that is supporting medium-term decisions. However, the accuracy of the method does not seem to meet the requirements for planning day-to-day operations. Nevertheless, these predictions may be of great help in resources planning and contribute to a better understanding of business activities. We strongly believe this this approach can support the exploration of strategies and provide a data-driven framework, leading to process optimization and cost reduction.

Several directions for future work can be enumerated. The workload prediction accuracy heavily depends on the RNN architecture. One suggestion is to train a model with a target window of more than one week (i.e., a sequence-to-sequence RNN) instead of re-injecting single predictions. Also, other algorithm families can be investigated to handle the different nature of the distributions. Furthermore, a more relevant score for evaluating the method should be considered. For instance, we could adapt the scoring function for including the quality of the predicted traces and their attributes (such as the timestamp, the duration, or the label).